\documentclass[fleqn,usenatbib]{mnras}

\usepackage{newtxtext,newtxmath}

\usepackage{txfonts}

\usepackage[T1]{fontenc}

\usepackage{amsmath}

\DeclareRobustCommand{\VAN}[3]{#2}
\let\VANthebibliography\thebibliography
\def\thebibliography{\DeclareRobustCommand{\VAN}[3]{##3}\VANthebibliography}

\usepackage{graphicx}	
\usepackage{amsmath}	

\usepackage{xcolor}

\definecolor{mycolor}{HTML}{007900}

\title[HSC-PDR3 extended PSFs]{The Hyper Suprime-Cam extended Point Spread Functions and applications}

\author[Garate-Nuñez et al.]{Lucía P. Garate-Nuñez,$^{1,2}$\thanks{E-mail: lucia.garatenunez@research.uwa.edu.au}
Aaron S. G. Robotham,$^{1,2}$
Sabine Bellstedt,$^{1}$ 
Luke J. M. Davies,$^{1}$\and
Cristina Martínez-Lombilla$^{2,3,4}$
\\
\\
$^{1}$ICRAR, The University of Western Australia, 35 Stirling Highway, Crawley, WA 6009, Australia\\
$^{2}$Australian Research Council Centre of Excellence for All-Sky Astrophysics in 3 Dimensions (ASTRO 3D), Stromlo, ACT 2611, Australia\\
$^{3}$School of Physics and Astronomy, Monash University, Clayton, Victoria 3800, Australia\\
$^{4}$School of Physics, University of New South Wales, Sydney, NSW 2052, Australia}

\date{Accepted 15 May 2024; Received 15 May 2024; in original form 22 Sept 2023}

\pubyear{2023}

\begin{document}
\newcommand{\SubItem}[1]{
    {\setlength\itemindent{15pt} \item[-] #1}
}
\label{firstpage}
\pagerange{\pageref{firstpage}--\pageref{lastpage}}
\maketitle

\begin{abstract}
We present extended point spread function (PSF) models for the Hyper Suprime-Cam Subaru Strategic Program Public Data Release 3 (HSC-SSP PDR3) in all \textit{g,r,i,Z} and \textit{Y}-bands. Due to its 8.2m primary mirror and long exposure periods, HSC combines deep images with wide-field coverage. Both properties make HSC one of the most suitable observing facilities for low surface brightness (LSB) studies, which are particularly sensitive to the PSF. By applying a median stacking technique of point-like sources with different brightness, we show how to construct the HSC-SSP PDR3 PSF models to an extent of R $\sim$ 5.6 arcmin. These models are appropriate for the HSC-PDR3 intermediate-state data which do not have applied the final aggressive background subtraction. The intermediate-state data is especially stored for users interested in large extended objects, where our new PSFs provide them with a crucial tool to characterise LSB properties at large angles. We demonstrate that our HSC PSFs behave reasonably in two scenarios. In the first one, we generate 2-D models of a bright star, showing no evidence of residual structures across the five bands. In the second scenario, we recreate the PSF-scattered light on mock images with special consideration of the effect of this additional flux on LSB measurements. We find that, despite the well-behaved nature of the HSC-PDR3 PSFs, there is a non-negligible impact on the faint light present in the mock images. This impact could lead to incorrect LSB measurements if a proper star subtraction is not applied.

\end{abstract}




\section{Introduction}
\label{intro}
Atmospheric conditions and the optics of the telescope, including its instruments and detectors, are the two main reasons why the light of a source is spread in astronomical images. To quantify this scatter, scientists have defined a function known as point spread function (PSF, \citealt{1999prop.book.....B}). The PSF describes the two-dimensional light distribution in the telescope focal plane produced by a source, where extensive studies have been performed to carefully characterise the imaging system.

The inner part of the PSF (within a few tens of arcseconds) is dominated by atmospheric turbulence \citep{1941DoSSR..30..301K} and is generally well represented by a Moffat function \citep{1969A&A.....3..455M, 1996PASP..108..699R}. An accurate determination of the inner part of the PSF is necessary to recover the real information of a point-like source. In the case of studying low surface brightness (LSB) emission, we also need to accurately determine the extended PSF profile to remove its faint scattered light which may contaminate flux measurements at large distances from a point-like source. The outer region of the PSF is less well-understood and was first measured by \citealt{1971PASP...83..199K}, who attributed an $r^{-2}$ behaviour to this part of the radial profile \citep{1958ApJ...128..465D}. Subsequent measurements mostly fitted the PSF wings using a power-law profile with power index ranges from 1.6 to 3 \citep{2005ApJ...618..195G, 2007ApJ...666..663B,2009PASP..121.1267S}.

In addition, the PSF light spreading in extended sources has also been widely studied. Initially, \citealt{1948AnAp...11..247D,  1953MNRAS.113..134D, 1958ApJ...128..465D} found that the effects of the PSF on large elliptical galaxies were of minor importance. However, more recent works have found that the measured amount of stellar light in the halos of galaxies could be significantly affected by the PSF \citep{1991ApJ...369...46U, 2014A&A...567A..97S, 2015A&A...577A.106S, 2016ApJ...823..123T, 2018ApJ...859...85T, 2019A&A...629A..12M}. As a consequence, PSF modeling will play a crucial role in two different areas: first, in recovering the spread of light from both point and extended sources, and second, in removing this contaminating light to robustly measure the emission from the own halos and nearby sources.

There are a couple of different methods that are mainly used in the literature to model a PSF. Star images can be combined to analytically extract the principal components of the PSF signature via the stacking method \citep{2012MNRAS.426.2300L, 2014MNRAS.443.1433D}. For example, \citealt{2008MNRAS.388.1521D} worked with data in the Hubble Ultra Deep Field (HUDF; \citealt{2006AJ....132.1729B}) of the Hubble Space Telescope (HST) and stacked Sloan Digital Sky Survey (SDSS; \citealt{2000AJ....120.1579Y}) images to estimate the effect of the PSF wings on edge-on galaxies, finding a significant impact on the measured stellar halos. 
\citealt{2013MNRAS.431.1121T} used the TinyTim HST PSF modeling software \citep{1995ASPC...77..349K} to generate the Hubble UDF PSF and conclude that scattered light from the extended PSF has a major impact on the surface brightness galaxy profiles. While in the case of \citealt{2020A&A...644A..42R}, the authors used SDSS Stripe 82 images to study the optical galactic cirrus by modeling the SDSS PSFs and correcting the scattered light of bright stars. These studies are essential in characterising the PSF from specific facilities and its impact on photometric measurements.


Telescopes that are highly optimized for low surface brightness have a true PSF that is well behaved and easy to model. The absence of reflective surfaces, anti-reflection coatings, an unobstructed pupil, and a fast focal ratio are the main characteristics of a telescope suitable for studying LSB sources \citep{2014PASP..126...55A}. 
\citealt{2009PASP..121.1267S} made an effort to mitigate the multiple internal reflections of bright point-like sources that complicate the estimation of the extended stellar PSF on the Case Western Reserve University’s Burrell Schmidt telescope located at the Kitt Peak National Observatory. These improvements helped to remove the excess light from extremely faint structures of the intra-cluster light (ICL) around the Virgo cluster \citep{2009ApJ...698.1879M, 2010ApJ...715..972J, 2010ApJ...720..569R}. More recently, in the case of the Dragonfly Telephoto Array \citep{2014PASP..126...55A}, due to its wide field of view and optical coatings designed to minimise reflections, the array is suitable for studying extended stellar halos of luminous nearby galaxies \citep{2016ApJ...830...62M, 2016ApJ...833..168M, 2018ApJ...855...78Z}. Since it is composed of telephoto lenses, the Dragonfly Telephoto Array is likely to have more stable PSF wings than reflecting telescopes, as the latter introduces more scatter \citep{2008SPIE.7012E..31N, 2022ApJ...925..219L}. For example, \citealt{2020MNRAS.495.4570M} indicated that the PSF effects are minimal at the Dragonfly galaxy outskirts. However, more stellar halo measurements are needed in order to analyse the effect of PSF wings.

Another competitive instrument for detecting low surface brightness emission is the Hyper Suprime-Cam (HSC; \citealt{2015ApJ...807...22M,2018PASJ...70S...1M, 2018PASJ...70S...2K,2018PASJ...70...66K}). Attached at the prime focus of the Subaru Telescope (Mauna Kea, Hawaii) and developed by the National Astronomical Observatory of Japan, HSC is one of the state-of-the-art cameras that provides a wide field of view and is also capable of reaching deep optical imaging in a short exposure time. The HSC team developed the Hyper Suprime-Cam Subaru Strategic Program (HSC-SSP; \citealt{2018PASJ...70S...8A}), a survey with a coverage of 1400 deg$^2$ in five different bands (\textit{griZY}) as well as four narrow filters. HSC-SSP is providing an unprecedented database for LSB studies. For example, \citealt{2018ApJ...857..104G} detected $\sim$ 800 low-surface brightness galaxies in the first 200 deg$^2$ of the Wide HSC-SSP layer, where approximately half of which are ultra-LSB with central surface
brightnesses in the \textit{g}-band of $\mu_0$> 24 mag arcsec$^{-2}$ \citep{2018ApJ...866..112G, 2021ApJ...920...72K}. \citealt{2021MNRAS.502.2419F} studied the ICL growth using a sample of 18 X-Ray clusters with deep HSC data in the \textit{i}-band between 0.1 < $\mathrm{z}$ < 0.5. Given its coverage, depth, and image quality (median seeing in all bands of $\sim$ 0.7 arcsec), this survey can be seen as a predecessor of the upcoming Vera C. Rubin Observatory Legacy Survey of Space and Time  (LSST; \citealt{2009arXiv0912.0201L}).

The HSC image processing pipeline \citep{2018PASJ...70S...5B,2019ASPC..523..521B} estimates the PSF using a restructured version of the
public PSFEx code \citep{2013ascl.soft01001B}. The approach of this code is to use detections that are likely to be point-like sources in the field to estimate the PSF from an exposure, where the PSF model can be constructed for any location on the image. However, \citealt{2017AAS...22922602M} found that the PSFEx model of the HSC PSF has super-resolution problems beyond a certain seeing threshold.

More recently, \citealt{2021ApJ...910...45M} constructed an HSC PSF with \textit{g} and \textit{i} observations from HSC-SSP to analyse the ICL of the Abell 85 cluster of galaxies. The authors followed the method outlined by {\cite{2020MNRAS.491.5317I}, which was originally designed for SDSS. In this method, the PSF construction process is divided into three parts, using median stacks of stars within a wide brightness range to build each region. \citealt{2021ApJ...910...45M} used four parts instead, where the two inner parts are made by stacking stars of faint magnitude and the two outer parts are made by fitting a power-law function to the brightest star profile in the field and extrapolating it up to R $\sim$ 7 arcmin. Two years later, \citealt{2023MNRAS.518.1195M} followed the same method to characterise the HSC PSF with observations from the HSC-SSP Public Data Release 2 \citep{2019PASJ...71..114A} and analyse the intra-group light (IGL) present in an intermediate-redshift Galaxy And Mass Assembly (GAMA, \citealt{2011MNRAS.413..971D}) group of galaxies.  In this case, the authors divided the PSF into three parts and applied a median stacking technique of stars in each part for the \textit{g,r,i}-bands up to an extent of R $\sim$ 5.6 arcmin in the \textit{g} and {r}-bands and R $\sim$ 4.2 arcmin in the \textit{i}-band. These PSF models are the result of the work presented in \citealt{2020RNAAS...4..124B}.

HSC-SSP PDR3 \citep{2022PASJ...74..247A} is the newest public data release from HSC, which is publicly accessible since August 2021. In comparison to both previous PDRs, this release increases the sky coverage to the required depths across all five filters. From PDR2 to PDR3, the partially observed area increased from 1114 deg$^2$ to 1470 deg$^2$, where the covered area at full depth ($\sim$ 26 mag at 5$\sigma$) increased from 305 deg$^2$ to 670 deg$^2$. In addition, due to a new global sky subtraction algorithm, the overall quality of the PDR3 data has also been improved and the extended wings of bright sources are better preserved. 
By building on advanced source extraction techniques and following the method from {\cite{2020MNRAS.491.5317I}, we characterise the HSC-SSP PDR3 PSFs in each of the $\textit{g,r,i,Z}$ and $\textit{Y}$ bands down to a radius of R $\sim$ 5.6 arcmin. The PSF image FITS files and the scripts to follow the PSFs construction are made available to the astronomical community.

The motivation to construct extended and well-characterised HSC-SSP PDR3 PSFs is to improve our future research on studying the intra-halo light (IHL, also known as intra-group or intra-cluster light) that is present in groups and clusters of galaxies with HSC observations. Following this idea, we present at the end of this paper the first step of our ongoing work, which consists in analysing the impact of the HSC-SSP PDR3 PSFs on low surface brightness sources present in mock images.

 The structure of the paper is as follows. In Section~\ref{Data} we discuss the data products used for this work. Section~\ref{PSF derivation} explains the steps we follow to construct the HSC-SSP PDR3 PSFs. Section~\ref{applications} is dedicated to testing the performance of the HSC-SSP PDR3 PSFs on two applications. We first use the PSFs to generate 2D models of an HSC-SSP PDR3 bright star in Section~\ref{star application}. Secondly, in Section~\ref{mock}, we use mock images to analyse the impact of the PSF-scattered flux on low surface brightness sources. We summarise the results of this work in Section~\ref{conclusions}.

We adopt throughout an $\mathrm{H_0}$ =  68.4 $ \, \mathrm{km}\,\mathrm{s}^{-1} \,\mathrm{Mpc}^{-1}$, $\mathrm{\Omega_\Lambda}$ = 0.699 and $\mathrm{\Omega_m}$ = 0.301 cosmological model, corresponding to the cosmology of \citealt{2020A&A...641A...6P}. All magnitudes are in the AB magnitude system and the HSC-SSP PDR3 zero point magnitude for all the five filters is m$_\text{zero}$ = 27 mag.

\section{The Hyper Suprime-Cam Subaru Strategic Program}
\label{Data}

Our data was collected by the Hyper Suprime-Cam, an optical-infrared imaging camera mounted on the prime focus of the 8.2-meter Subaru Telescope. It is operated by the National Astronomical Observatory of Japan (NAOJ) and is located at the Mauna Kea Observatory in Hawaii. The large focal plane is paved with a mosaic of 116 charge-couple devices (CCDs) detector formatting 2k × 4k, covering a field of view of $\sim$ 1.7 deg$^2$ with a pixel scale of $\sim$ 0.168 arcseconds. 

In this work, we use data from the Hyper Suprime-Cam Subaru Strategic Program \citep{2018PASJ...70S...8A}. In particular, the images are from the last public data release of the HSC-SSP Survey, HSC-SSP PDR3 \citep{2022PASJ...74..247A}. We use the Gaia Data Release 3 catalogue \citep{2022arXiv220800211G} to find the positions of the stars that are used to obtain the empirical PSF models.  

HSC-SSP consists of three layers (Wide, Deep \& UltraDeep) and uses five broad-band (\textit{griZY}) and four narrow-band filters. Each of the layers has a different sky coverage and depth: 1400 deg$^2$ (r $\sim$ 26 mag),  27 deg$^2$  (r $\sim$ 27 mag), and 3.5 deg$^2$ (r $\sim$ 28 mag) for the Wide, Deep, and UltraDeep layers respectively.

\subsection{PDR3}
To date, the HSC-SSP has three public releases: PDR1  \citep{2018PASJ...70S...8A}, PDR2 \citep{2019PASJ...71..114A} and PDR3 \citep{2022PASJ...74..247A}. Every new PDR is a major update in terms of depth and area in comparison to their predecessors. In addition, a significant number of improvements in the data processing HSC pipeline have taken place since HSC-SSP PDR1 data was released, with careful consideration of the sky estimation capabilities \citep{2023MNRAS.520.2484K}. The sky background estimation remains one of the challenges in low surface brightness imaging and is one of the major sources of systematics \citep{2016MNRAS.456.1359F,2017ApJ...834...16M,2023ApJ...953....7L}.

PDR1 oversubtracted the sky around the extended wings of stars and bright galaxies. This issue was mitigated by a global sky subtraction introduced in PDR2 (see Figure 5 of \citealt{2019PASJ...71..114A}), which subtracted the background using superpixels of 1k $\times$ 1k pixels ($\sim$ 168" on a side). However, this presented a few disadvantages, some of them introduced by the fact that the size of the superpixel was smaller than the size of a single CCD (11.5' {$\times$} 5.7'). In PDR3, the new algorithm consists of gridding a visit image into superpixels of 8k $\times$ 8k ($\sim$ 23'\ on a side). As the size of the superpixel is larger than the size of a CCD, the sky frame is now less biased by avoiding discontinuities at CCD edges. In addition, the PDR3 global sky subtraction algorithm does a better job at preserving the extended halos of large objects. At this stage, background fluctuations at small-scale were present in PDR3 frames and made the authors add a local sky subtraction using superpixels of 256 $\times$ 256 pixels ($\sim$ 43" on a side). By properly masking the sources and applying an iterative process, the authors re-estimated the sky level locally, which mitigated this issue. However, after the 8k $\times$ 8k and 256 $\times$ 256 sky subtractions, false detections and measurement failures around the extended wings of bright stars and galaxies were still present. To solve this problem, the authors added a second local sky subtraction with a superpixel size of $\sim$ 21.5" on a side. This final subtraction compensates for the detection efficiency, but it also overestimates the sky around bright sources, affecting the faint diffuse light that surrounds these sources.

Following this idea, there are two types of coadd images in PDR3: the first with the three sky subtractions applied, and the second with only the first two sky subtractions applied. As we are interested in studying the intra-halo light surrounding extended objects, we need the well-preserved wings of the HSC bright star images to construct extended PSFs. For this reason, we decided to work with the latter type of coadded image. This intermediate-stage data is not publicly available on the HSC-SSP website via the Data Archive System (DAS) search tool and we contacted the HSC software team to download it. However, it is publicly available via the Image Cutout tool under the "coadd/bg" file type option.

In conclusion, our HSC-SSP PDR3 PSF models are appropriate for data at the intermediate stage where the final local sky subtraction has not been applied. This data is useful for low surface brightness studies where extended objects might be affected by aggressive background subtractions.

\section{HSC-SSP PDR3 PSF derivation}
\label{PSF derivation}

We apply an empirical method to construct the PSF for the HSC-SSP PDR3 in each of the five $\textit{g,r,i,Z}$ and $\textit{Y}$ bands. In this method, we stack stars of different brightness using the median operator to construct the different radial ranges of the PSF. We have made available the PSF construction scripts written in the open-source R language at \,\url{https://github.com/luciagarate/HSC\_PSFs}. This consists of three scripts: the first describes the selection of the star images for each PSF region ("1$\_$StarSelection.R", Section~\ref{selection}), the second describes the subsequent stacking of the selected stars ("2$\_$Stacking.R", Section~\ref{stacking}),  and the third describes the combination of the different regions to construct a PSF per HSC filter ("3$\_$Combining.R", Section~\ref{combination}). The HSC-SSP PDR3 PSFs in all five bands are also available publicly as FITS files.

\subsection{Star selection}
\label{selection}
We follow a similar approach to \citealt{2020MNRAS.491.5317I}, where the authors originally constructed a PSF for the SDSS.  
This method consists of constructing different parts of the PSF by the median-clipping stacking of stars within a wide brightness range. In particular, the authors divide the PSF construction into three parts. Bright and saturated stars are used to characterise the outer part (or wings) of the PSF, stars of intermediate brightness are used for the middle part and the core is characterised by non-saturated faint stars. As the central pixels of bright stars are highly affected by the saturation of the CCD dynamical range, the authors only use the outer pixels of these images to create the faint PSF wings. However, the central pixels of fainter stars are unaffected by saturation and can be used to construct the PSF core. 

Based on our experimentation, in this work, we divide the PSF into four radial regions: outer, middle, inner, and core, where each part has a different brightness range. These four selected magnitude ranges are: 

\begin{itemize}
    \item Outer: mag$_{\star, g}$ < 8 ($\sim$ 1200 stars)
    \item Middle: 11 < mag$_{\star, g}$ < 11.5 ($\sim$ 1700 stars)
    \item Inner: 14 < mag$_{\star, g}$ < 14.1 ($\sim$ 2000 stars)
    \item Core: 18 < mag$_{\star, g}$ < 18.02 ($\sim$ 2300 stars)
\end{itemize}

As previously mentioned, we use the Gaia Data Release 3 catalogue \citep{2022arXiv220800211G} to identify stars that satisfy these magnitude requirements in the $\textit{g}$-band data of the catalogue. An approximation of the number of stars that were selected to construct the different parts of the PSF is indicated above. For the middle, inner, and core regions, we make the selection of stars in the overlapping regions between HSC-SSP PDR3 and the Galaxy And Mass Assembly (GAMA, \citealt{2011MNRAS.413..971D}) survey. This survey overlaps with HSC-SSP PDR3 in the GAMA-fields\footnote[1]{The HSC-SSP survey selection of targets was originally designed to overlap with other multi-wavelength data, including near/mid-infrared imaging surveys such as GAMA, to maximise the scientific synergy with HSC.} G09, G12 \& G15, and partially overlaps in G02, with a total overlapping area of $\sim$ 200 deg$^2$. The quantity of stars in this overlapped region for the middle, inner, and core PSF regions is adequate for this work. Only for the outer region, we make the star selection using all HSC-SSP PDR3 Wide data, as the number of bright stars in the HSC-GAMA overlapped region is limited. 

We then also apply an additional constraint that rejects the stars that are not within one degree of radial distance from the centre of any HSC patch \footnote[2]{The HSC data is stored in tracts of 9 $\times$ 9 patches. Each tract is $\sim$ 1.7 deg$^2$
and each patch is $\sim$ 11.9 arcmin$^2$ ($\sim$ 4200 pix$^2$). There is an overlap of 1 arcmin between the two adjacent tracts and 34 arcsec between patches.}. By doing this, we avoid distortion problems around the borders of the patches.

The size of each HSC cutout used to construct the outer PSF is 4k $\times$ 4k pix${^2}$, meaning that after applying the stacking and combination process (see below for full details), the resulting PSFs extend to a radius of R = 2000 pixels ($\sim$ 5.6 arcmin) in all the five bands. As the rest of the selected stars are used to construct the inner parts of the PSF, the cutout sizes are smaller: R = 1000 pixels for the middle stack, R = 500 pixels for the inner stack, and R = 200 pixels for the core one.

\subsection{Stacking}
\label{stacking}

After selecting stars and generating cutout HSC images per filter, we proceed to construct the different parts of the HSC PDR3 PSF. Section~\ref{outer} explains how we build the outer part of the PSF. Section~\ref{middle and inner} describes the steps we followed to build the middle and inner parts of the PFS, and Section~\ref{core} contains the details about the construction of the core.

\subsubsection{Outer part of the PSF}
\label{outer}
A total of $\sim$ 1200 stars are used to construct the PSF wings. This process consists of applying a median stack to all the selected stars by making use of the warping and stacking package \textsc{ProPane}\footnote[3]{\url{https://github.com/asgr/ProPane}} \citep{2024MNRAS.528.5046R}.

To mask the background sources in the star images we use the \textsc{ProFound}\footnote[4]{\url{https://github.com/asgr/ProFound}} \citep{2018MNRAS.476.3137R} source finding and photometry analysis package. \textsc{ProFound} detects sources using dilated segments (ishophotal outlines) of arbitrary shape (rather than elliptical
apertures) and measures statistics like flux, size, and ellipticity.

In particular, we run the \texttt{profoundImDiff} function on the star image, which creates another image that is the result of the original minus a smoothed version (in this version the borders of the segmentation maps have no abrupt changes). The parameter that establishes the standard deviation of the blur is \texttt{sigma}. We then apply the \texttt{profoundDilate} function to the resulting image, which dilates the segmentation maps by typically 30\%. During the dilation process, a watershed de-blending approach is taken, where the segments are not allowed to overlap. The dilation is important to make sure that the background sources are completely masked. 
The \texttt{size} parameter controls the width/diameter of the dilation kernel in pixels. We then mask all of the pixels inside each segment. The selected \texttt{sigma} and \texttt{size} values for each region and filter are indicated in the script "1$\_$StarSelection.R". We take the extra precaution of leaving unmasked the horizontal feature (bleeding) observed in bright HSC star images across all bands due to CCD saturation and also the vertical feature only present in the \textit{Y}-band (see Appendix~\ref{appendix} for further details). In addition, for star images in the \textit{g,i,Z,Y}-bands, we leave unmasked the region within R = 200 pix from the centre.

Before stacking, we normalise each image by the value of the median flux contained in an annulus situated between 400 and 410 pixels from the centre of the bright star. The position of the annulus was chosen to be as close to the centre as possible but avoiding the central saturated region of each image. The width of the annulus was selected in order to get a good signal-to-noise ratio (S/N). The upper Fig.~\ref{mask} shows an image of a star used to construct the outer part of the PSF with the corresponding masks. The black circles show the normalisation annulus. As we are calculating the median flux inside the annulus, our masking approach is not very aggressive. The lower plot shows a histogram of the pixel flux values within the annulus of the upper image, where the vertical red line indicates the median value used to normalise the flux.  

Images with a total masked area larger than 80\% are discarded for the stacking process. Finally, we proceed to make a median stack of the masked normalised images with the \texttt{propaneStackFlatFunc} function, which stacks already aligned (flat) images. The same normalisation process is applied in all the HSC filters.

\begin{figure}
\includegraphics[width=1\linewidth]{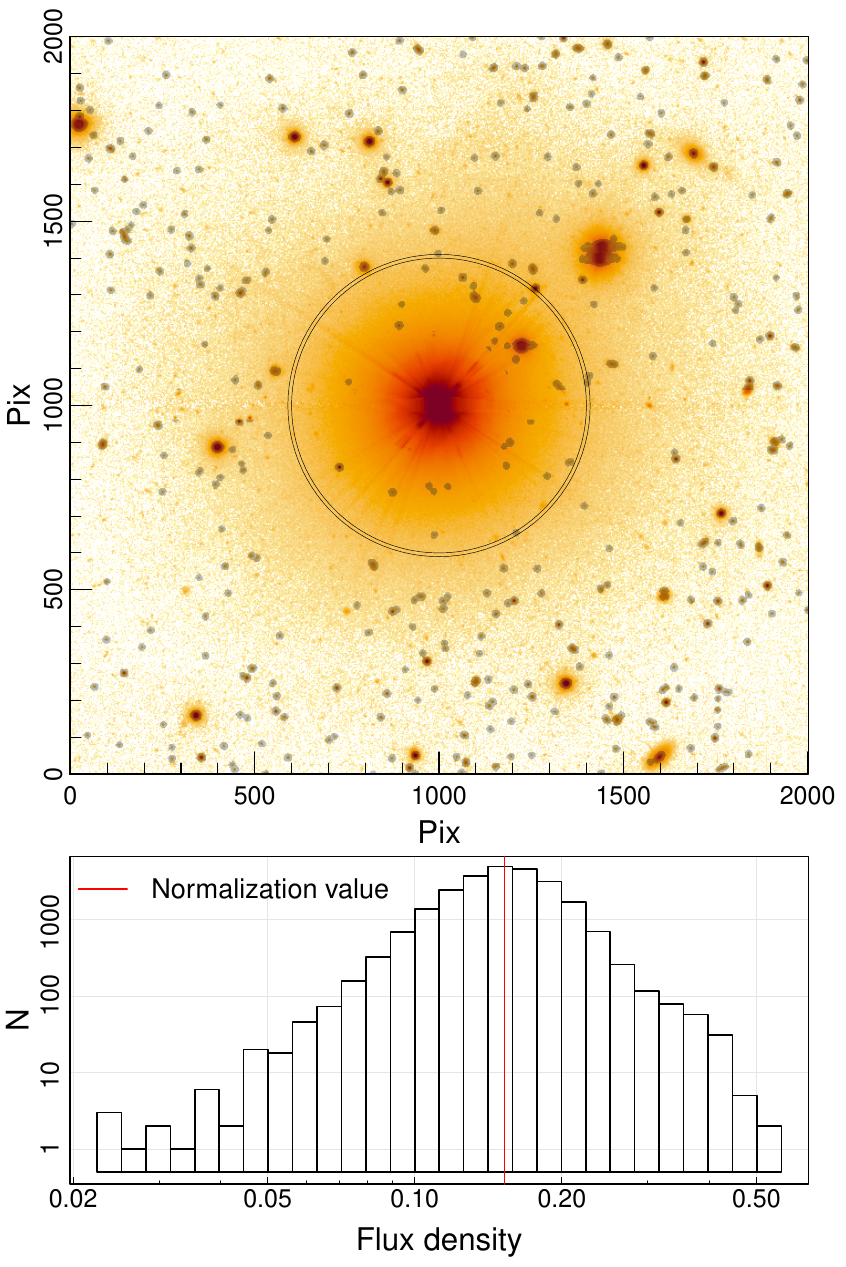}
    \caption{Upper: image of a star used to construct the PSF outer part where the \textsc{ProFound} masks are indicated by the semi-transparent gray spots. The normalisation annulus used to stack all the images is delimited by the two black thin circles. Lower: histogram of the flux values of all the pixels within the annulus from the upper image, where the red vertical line indicates the median value of this distribution. }
    \label{mask}
\end{figure}

The leftmost panel of Fig.~\ref{PSF steps norm} shows the stack of the stars used to construct the PSF outer part with the selected annulus for the normalisation.

\subsubsection{Middle and Inner parts of the PSF}
\label{middle and inner}
The methodology to construct the middle part of the PSF is similar to that in Sec.~\ref{outer}. The main difference is in the normalisation step, where we select the annulus of inner radial distance 200 pixels from the centre and width 20 pixels. In addition, the size of each image used is 1k $\times$ 1k pix${^2}$, and images, where the masked area is more than 50\%, are not considered for the stacking process.

In the case of the inner PSF construction step, the size of each image is 500 $\times$ 500 pix${^2}$, and the normalisation annulus has an inner radius of 100 pixels and an outer radius of 140 pixels. To stack, we only select the images of stars that have less than 2\% of pixels masked. We then proceed to apply the median stack with the masked images. 

 The middle and inner stacks with their respective selected annuli are shown on the second and third panels of Fig.~\ref{PSF steps norm}.

\subsubsection{Core of the PSF}
\label{core}
 The inner radius of the annulus for the normalisation step in the core stack is 10 and the outer is 15 pixels, where the size of each image is 200 $\times$ 200 pix${^2}$. 
 In this case, the masks are calculated using the main function of \textsc{ProFound}, \texttt{profoundProFound}. This function offers great flexibility to create large segments around background sources, which is important due to the faintness of the central star. Images that were selected for the stacking process have less than 20\% of pixels masked. We also discard the stars that do not have the brightest pixel in the centre of the image. Due to GAIA coordinate errors, the brightest pixel of the HSC image can be uncentred and we only want stars with low proper motion. As in the other three cases, we calculate the median stack with the selected masked images.
 
 The rightmost panel of Fig.~\ref{PSF steps norm} shows the core stack with the selected annulus to normalise the images.

\begin{figure*}
\centering  
 \includegraphics[width=1\linewidth,trim=2.5 8 12 8,clip]{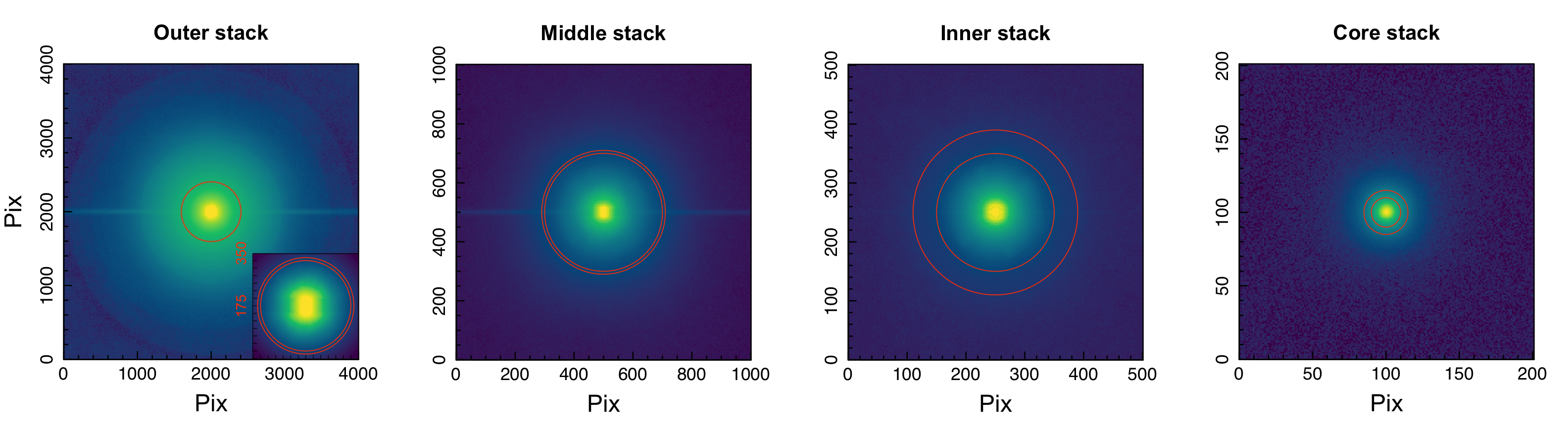}

    \caption{First panel shows the outer stack (with zoomed-in central region of R $<$ 350 pix), the second panel shows the middle stack, and the third and fourth panels are the inner and core stacks, all with their respective chosen annuli to normalise each star image indicated by thin red circles. The sizes of the respective stacks are: 4k $\times$ 4k pix${^2}$, 1k $\times$ 1k pix${^2}$, 500 $\times$ 500 pix${^2}$ and 200 $\times$ 200 pix${^2}$. All the images are in the $\textit{g}$-band.}
    \label{PSF steps norm}
\end{figure*}

\subsection{Combination}
\label{combination}
The four separate median stacks are combined to obtain the final HSC PSF per filter. 

We begin by combining the outer and middle stacks. To combine both parts, we first normalise them separately by calculating the median flux within an annulus of an inner radius of 150 pixels and an outer radius of 160 pixels in each of both stacks. We then replace inside a circle of radius 150 pix in the outer stack with pixels from the middle stack, henceforth outer + middle. Similarly, we combine the outer + middle with the inner stack, and finally the outer + middle + inner with the core stack. The selected annuli to normalise the different parts are indicated in Table~\ref{combination table}, as well as the chosen radii to combine them.

\begin{table*}
\centering
\begin{tabular}{|c|c|c|}
 \textbf{} &\textbf{Normalisation annulus} & \textbf{Combination radius}                                                                                                                                                        \\ \hline
                     Outer   &       160-150         &   150                                    \\ \hline
Middle  &
160-150 & 
\\ \hline
                     Outer + Middle   &       70-60         &  60                                                                                                                                                      \\ \hline 
                      Inner   &       70-60         &   
                      \\ \hline 
                       Outer + Middle + Inner  &       30- 5        &  20
                      \\ \hline 
                      Core   &       30-5          &  \\ \hline 
\end{tabular}
\caption{Annuli apertures used for the normalisation of the different PSF parts in the combination process, along with the radii used for the combination of these different parts.}
\label{combination table}
\end{table*}   

The values of the annuli to normalise each image and radii to combine the regions were chosen by analysing the radial distribution of each stack, where we look for the brightest inner radius that has a good S/N in both parts and where the merging profiles agree well. We show the step-by-step radial distributions\footnote[5]{In this paper, all the radial distributions show the flux values of every pixel of the respective image as a function of their distance to the center. 
It is important to note that no pixel values are excluded, and no statistical methods are applied to present these radial distributions.} of the combined stacks in the left panel of Fig.~\ref{PSF profiles}, and the radial distribution of the single PSF stacks in the right panel . 
The salmon line in the first panel is the radial distribution of the outer stack, whereas the blue line represents the distribution of the outer and middle stack combination. The radius of this junction (150 pixels) is indicated with the right dashed grey line. Following this idea, the green line represents the combination of the three outer parts, and the purple one is the final combination. The 60 and 20-pixel junction radii are indicated by the middle and left dashed grey lines respectively. All the selected annuli to normalise and radii to replace are the same across the five HSC filters. As mentioned above, we show the radial distributions of the median single PSF stacks in the right panel of Fig.~\ref{PSF profiles}, where each point represents each pixel of the selected PSF region. The salmon points represent the outer region of the PSF, the blue and green points are the pixels from the middle and inner PSF regions parts respectively, whereas the purple points represent the core of the PSF. We also show the flux error associated with each of the four median stacks in pale blue. The error per pixel associated to the median stack is estimated by calculating the 1$\sigma$-quantile (half of the range containing the central 68$\%$ of data) of each stacked pixel multiplied by 1.253 and divided by the square root of the number of contributing pixels. 

The spread in the flux pixels values of the radial distributions at R $>$ 1000 pix is a factor of $\sim$ 5 larger than the median of the stacking error associated with each individual pixel. We attribute this large scatter to systematic uncertainties in the background subtraction process.

 Fig.~\ref{PSF steps} shows the images of the above-mentioned combinations for the different parts: the first panel is the outer stack,  the combination of the outer and middle parts is shown in the second panel, the third panel shows the combination of the two outer parts with the inner one and the final combination (outer + middle + inner + core) is shown in the fourth panel. Each panel has delimited with red circles the annuli apertures used for the normalisation of the different PSF parts in the combination process and the radii used for the combination of these different parts (see Table~\ref{combination table}). The rectangular black box in each panel is a zoomed-in view of the central region. We use the same color scale with identical minimum and maximum values to map each panel.

After combining the four median stacks, we obtain the HSC PSFs with a radial extension of R $\sim$ 5.6 arcmin. We then proceed to apply the next two final details to each of the PSFs in order to obtain the final HSC PSFs per band:

\begin{itemize}

    \item Assuming the HSC PDR3 PSFs are rotationally symmetric, we apply horizontal, vertical, and diagonal reflections to obtain a symmetric 2-D PSF. To do this, we create a new image with the same dimension as the original PSF. For each pixel in the new image, we sum its original value with the values of the three reflected pixels values: the horizontally reflected pixel, the vertically reflected, and the diagonally reflected one. However, in optical systems with a wide field of view, significant PSF spatial variations might appear. We release both PSF versions (before and after symmetrisation), where the user should choose which one to use depending on their science goals. We make a comparison between the symmetric and asymmetric HSC-PDR3 PSF in Appendix~\ref{comparison symm}.

    \item We then normalise the flux of each PSF image by making the sum of all the pixel values equal to one.
\end{itemize}

\begin{figure*}
\centering  
 \includegraphics[width=1\linewidth]{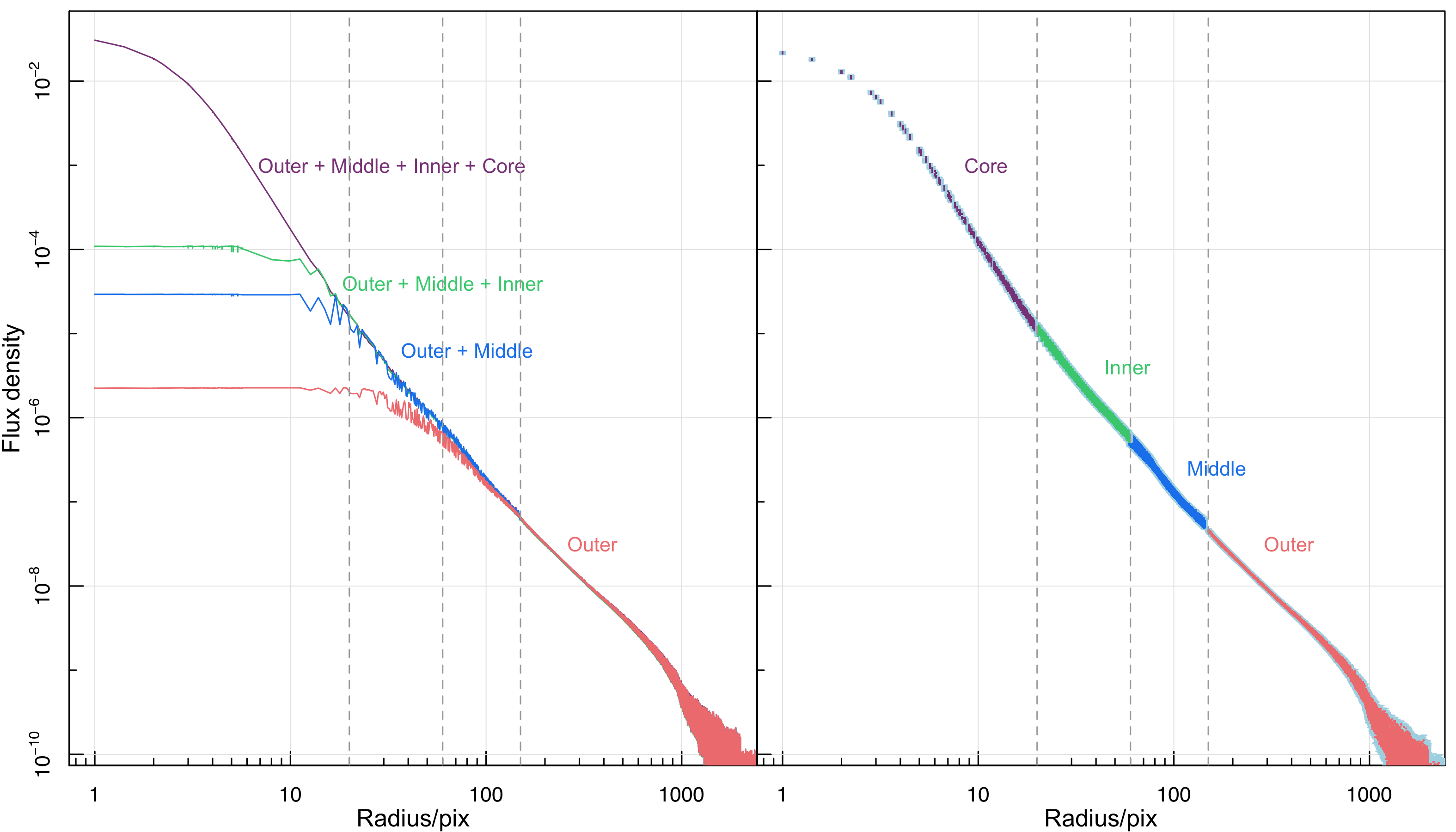} 
    \caption{First panel: radial distributions of the outer stack (salmon), the combination of the outer and middle stacks (blue), the combination of the outer, middle, and inner stacks (green), and the combination of the outer, middle, inner and core stacks (purple), which is the final HSC $\textit{g}$-band PSF. Second panel: contribution of each individual stack to the final radial distribution, where the outer stack is represented by the salmon points, the middle stack by the blue points, the inner stack is indicated by the green points, and the core stack by the purple points. We also show the flux median stack error per pixel in pale blue, where this error is calculated as the 1$\sigma$-quantile multiplied by 1.253 and divided by the square-root of the number of pixels that contributed to each median stack. In both panels, the right dashed grey line indicates the radius where we make the first combination (150 pix), the middle one indicates the radius where we join together the outer + middle with the inner stack (60 pix) and the left one indicates the radius where this combination and the core stack are combined (20 pix). All the distributions are made with observations in the $\textit{g}$-band. As previously mentioned, a radial distribution shows the flux distribution of all the pixels of the respective image as a function of their distance to the center.}
    \label{PSF profiles}
\end{figure*}

\begin{figure*}
\centering  
 \includegraphics[width=1\linewidth]{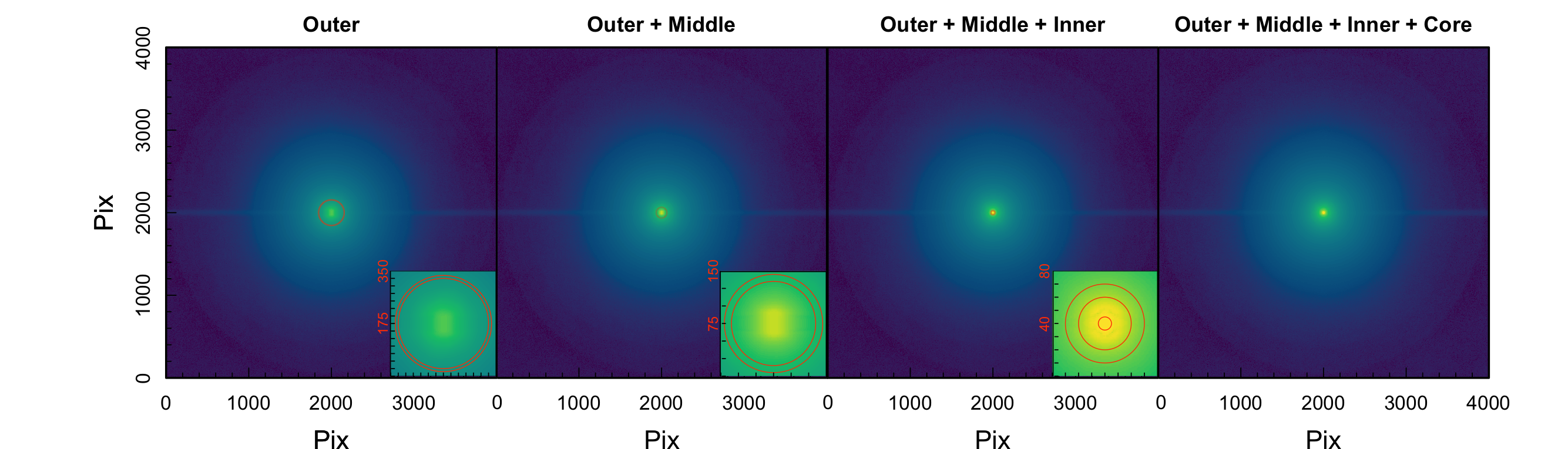}

    \caption{Left panel: outer PSF stack, where the annulus that was selected to normalise both outer and middle stacks is represented by the two thin red circles. The inner radius (150 pix) delimits the pixels that we replace with the middle stack. Second panel: outer + middle combination with the annulus of outer radius 70 pix and inner one 60 pix, where we replace the inner stack. Third panel: outer + middle + inner combination with the normalisation annulus situated between 30 (outer) and 5 (inner) pixels from the centre of the image.  We replace with the core stack inside the 20 pixel red circle delimited by the intermediate radius. Right panel: HSC $\textit{g}$-band PSF before applying the final details (normalisation + symmetrisation). For further clarity, zoomed-in views are indicated with a rectangular black box in each panel. We use the same color scale with fixed upper and lower limits across the four panels.}
    \label{PSF steps}
\end{figure*}

In the left panel of Fig.~\ref{final psfs} we show the final $\textit{g}$-band HSC PSF image up to R $\sim$ 5.6 arcmin (2000 pixels). In the right panel, we show its flux radial distribution, where the horizontal dashed line indicates the maximum flux value of the PSF divided by two ($\mathrm{f_{max}}$/2), and the vertical one indicates the full width at half maximum (FWHM) of the PSF also divided by two. 
In the $\textit{g}$-band, the FWHM is 0.68 arcsec ($\sim$ 4 pixels). The difference in flux between the peak and the minimum value is $\Delta$flux $\sim$ 10$^8$, which is equivalent to $\Delta$mag $\sim$ 20.

In the left panel of Fig.~\ref{final psfs} we can easily identify some of the CCD artefacts (mainly more remarkable around bright stars) described in \citealt{2022PASJ...74..247A}. These artefacts are unwanted or unintended effects that occur in CCD images (noise, blooming, hot pixels, dead pixels, etc) where some of them are identified and interpolated over in the CCD processing, but some artefacts may still be present in the processed images.   

The bleeding (or blooming) due to CCD saturation is the horizontal feature that is present in all five bands. We see this feature not only in the outer stack but also when stacking the stars from the middle region and even when stacking the inner region in some of the bands. The next observable artefact is the ghost, produced by internal reflections in the HSC. Due to the multiple paths of reflections, we see ghost features at different angular sizes. Another observable artefact in Fig.~\ref{final psfs} is the smooth halo component around the centre of the PSF, which is also present in all the stacks. We show the HSC-SSP PDR3 PSFs in the other four bands in Appendix~\ref{appendix}, where more artefacts are visible.

 \begin{figure*}
\centering  
\includegraphics[width=1\linewidth]{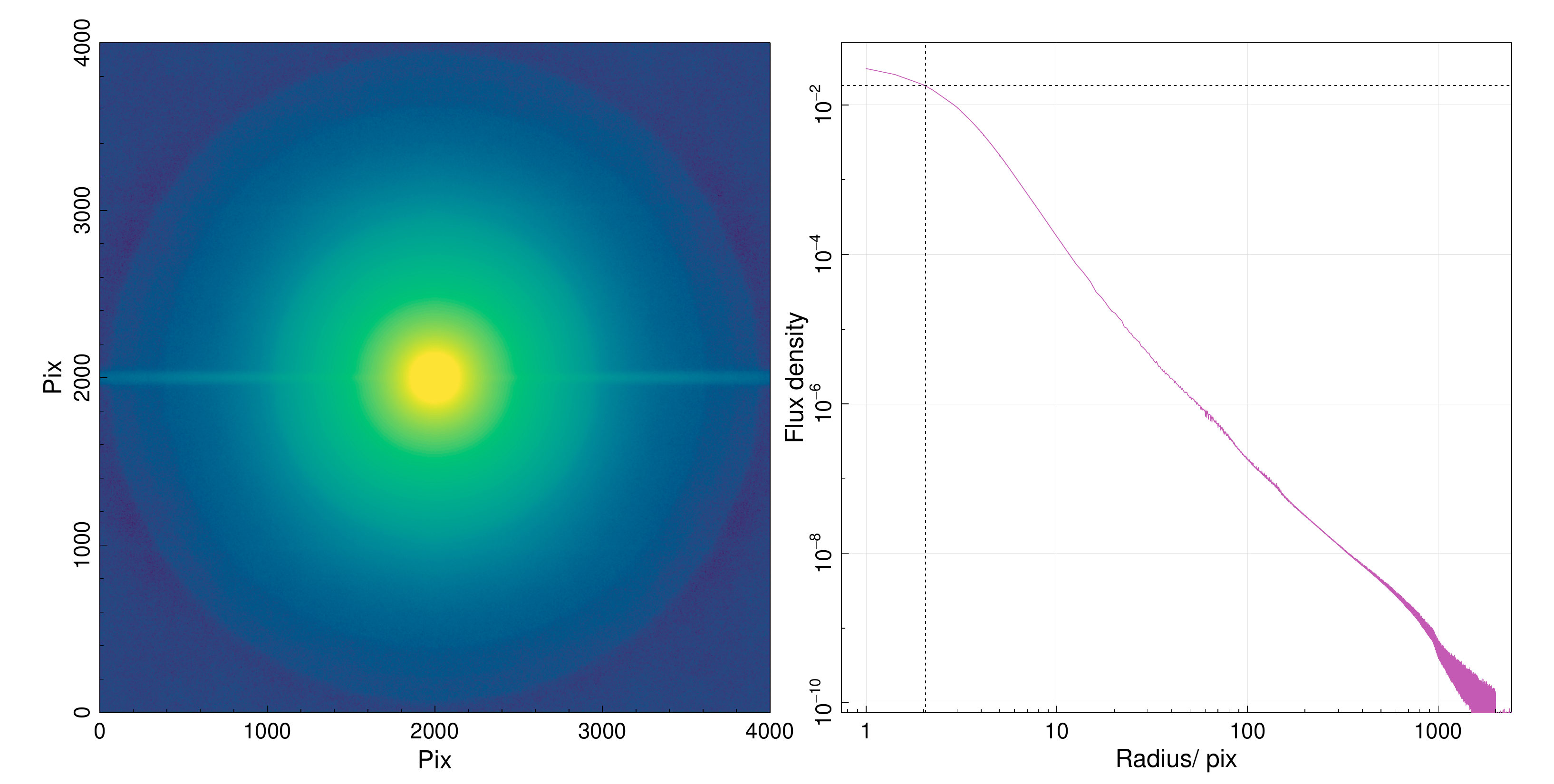} 

    \caption{HSC-SSP PDR3 PSF in the \textit{g}-band. The left panel is the image of the PSF and the right panel is the radial distribution. The PSF is normalised to have a total flux equal to one. The horizontal dashed line indicates the $\mathrm{f_{max}}$/2 of the profile, where $\mathrm{f_{max}}$ is the maximum flux value, and the FWHM/2 ($\sim$ 0.34 arcsec) is specified by the vertical dashed line.}
    \label{final psfs}
\end{figure*}

Finally, in Fig.~\ref{PSF comp} we compare our \textit{r}-band HSC-SSP PDR3 PSF flux profile (dark blue) + extrapolated dashed red line up to 35 arcmin versus the \textit{r}-band PSF flux profiles of the 0.9 m Burrell Schmidt telescope (green, \citealt{2009PASP..121.1267S}) and the Dragonfly Telephoto Array (pink). The solid pink Dragonfly PSF \citep{2014PASP..126...55A} and the Burrell Schmidt PSF show the \textit{r}-band AB mag arcsec$^{-2}$ surface brightness profile of Vega (see \citealt{2014PASP..126...55A} for details). The former was based on direct profile extraction, by stitching several subprofiles with a range of integration times, extending out to a radius of R = 1$^{\circ}$. On the other hand, the dashed pink Dragonfly PSF \citep{2022ApJ...925..219L} is a representative model PSF reconstructed from a typical deep Dragonfly image. The four profiles are normalised to have a flux equal to one within the range of 0.1 and 5 arcmin and the minimum radius plotted is 0.1 arcmin. In all the four PSFs, more than 90\% of the total flux is contained at R $<$ 0.1 arcmin \citep{2014PASP..126...55A}, with our PSF accounting for 97.2\% of
the total light within R $<$ 0.1'. This means that less than $\sim$ 3 \% of the total flux is contained in the extended wings of the HSC-SSP PDR3 PSF. Both the Burrell Schmidt telescope and Dragonfly Telephoto Array are highly optimized for low surface brightness imaging and have well-behaved PSFs with low power distributed in the far wings \citep{2022ApJ...925..219L}. In comparison to them, we find the performance of our HSC-PDR3 PSF to be remarkably good.

\begin{figure}
\centering  
 \includegraphics[width=1\linewidth]{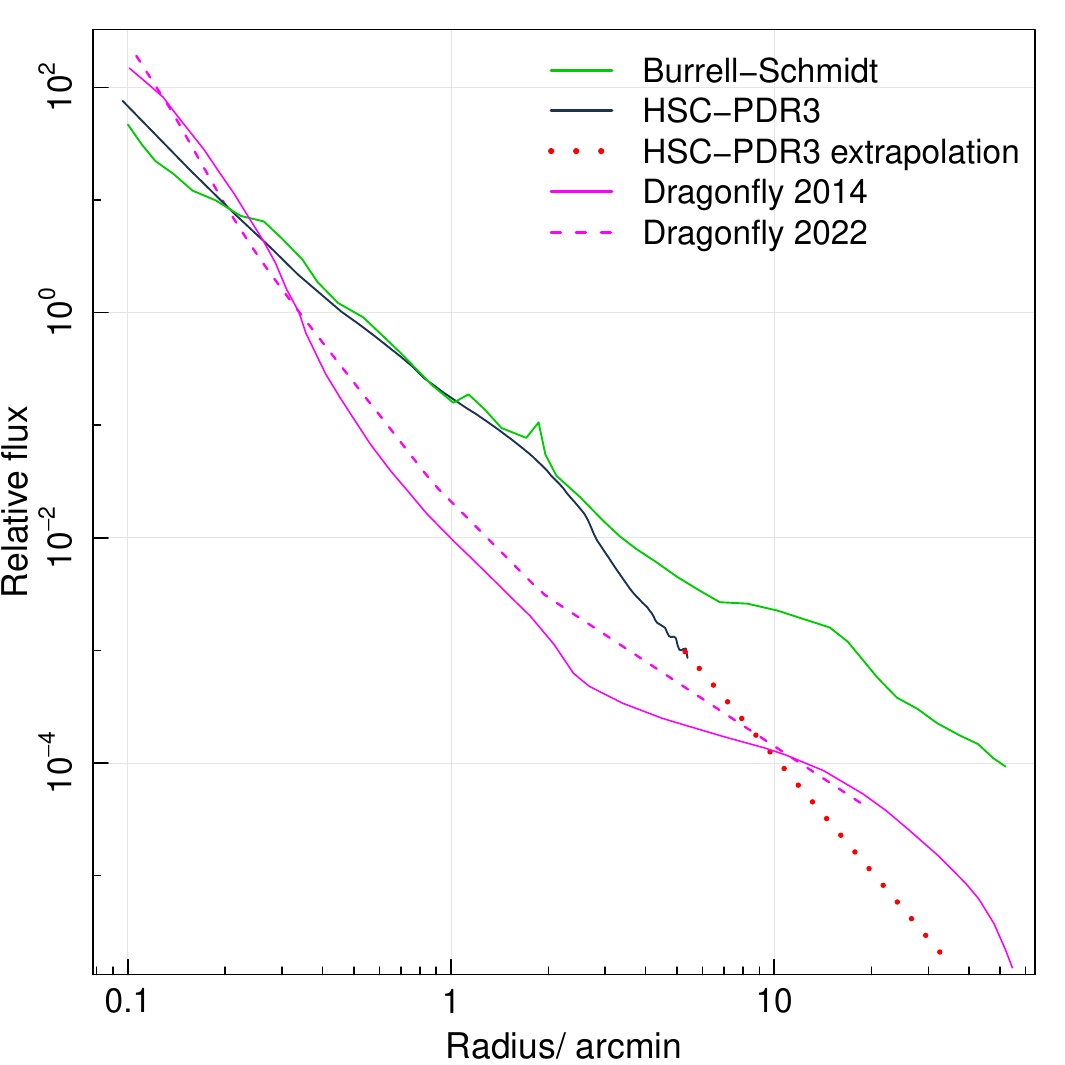}

    \caption{Comparison of the \textit{r}-band PSF radial profiles of Burrell Schmidt telescope (green), Dragonfly Telephoto Array 2014 (solid pink), Dragonfly Telephoto Array 2021 (dashed pink) and our HSC-PDR3 model (dark blue) with an extrapolation up to 35 arcmin shown as the red dashed line. Each of the profiles is normalised so that the flux between 0.1 arcmin and 5 arcmin is one.}
    \label{PSF comp}
\end{figure}

\section{PSF applications} 
\label{applications}
In this section, we show two examples of our new HSC-SSP PDR3 PSFs performance when used in scientific applications. The first application consists of using our empirical PSFs to generate a two-dimensional model of an HSC-PDR3 star image. 
Additionally, in the second application we generate mock images and use our HSC-PSF models to measure the impact of the PSF-scattered light on the faint end of the surface brightness distributions.

\subsection{Star application}
\label{star application}
In the first application, we use the HSC-SSP PDR3 PSFs to fit a two-dimensional model per band to HSC-PDR3 observations of a bright star. The images of this star in the five different HSC filters are shown in the first row of Fig.~\ref{PSF star application}. The next row shows the masked version of these images, where we use the \textsc{ProFound} package to mask the background sources and the cores of the stars. The central part of each image is masked up to a radius where the pixels are no longer saturated. This masking limit can easily be determined by plotting the star radial distribution. The reason behind avoiding the saturated pixel values is that they introduce fake flux measurements and could have a dramatic effect on the subsequent star profiling. Once we have correctly masked the images, we proceed to make the 2-D  star profile modeling per band with \textsc{ProFit}\footnote[6]{\url{https://github.com/ICRAR/ProFit}} \citep{2017MNRAS.466.1513R}. 
\textsc{ProFit} is a Bayesian 2-D galaxy profiling tool, that by default offers a set of popular profiles (Sérsic, Moffat, broken-exponential, etc) to generate an image model of a model light profile. In particular, we use the point spread function profile template, which takes the point-like source image, convolves it with the PSF provided, and effectively renormalises it to match the magnitude of the star image. The resulting 2-D star models per band are shown in the third row of Fig.~\ref{PSF star application}, and the fitted star magnitudes are indicated in the titles of each column. In the fourth panel, we present the residuals obtained by subtracting the models from the masked observations, revealing some expected residuals around the core of the stars as a consequence of the differences between the individual star and the PSF averaged model. However, when plotting the residuals divided by the corresponding image (see fifth row), we see that the percentage difference is smoothed around the central regions. In the sixth row we also show the difference between the image (with background sources unmasked) and the model, but with the same color scale that is used to plot the star images in the 1\textsuperscript{st} row, showing the performance of our HSC-PDR3 PSF models in a star subtracted image. Finally, in the last row, we show the radial distribution of the star up to a radius of R $\sim$ 1500 pixels in light blue, the masked star radial distribution in dark blue, and the \textsc{ProFit} model radial distribution in red.  The dashed black lines indicate the radii where the core of each star is saturated. We observe that the point-source models that were constructed with our empirical HSC-SSP PDR3 PSF images reproduce well the data across all five filters, including the vertical spikes in the \textit{Y}-band star image. The small peaks at low radii present in all the red profiles is a discretisation artefact of \textsc{ProFit}. We also see that the noise of the \textsc{ProFit} models is reduced compared to the noise of the individual star images as a consequence of the PSF stacking technique.

\begin{figure*}
\centering  
 \includegraphics[width=.88\linewidth]{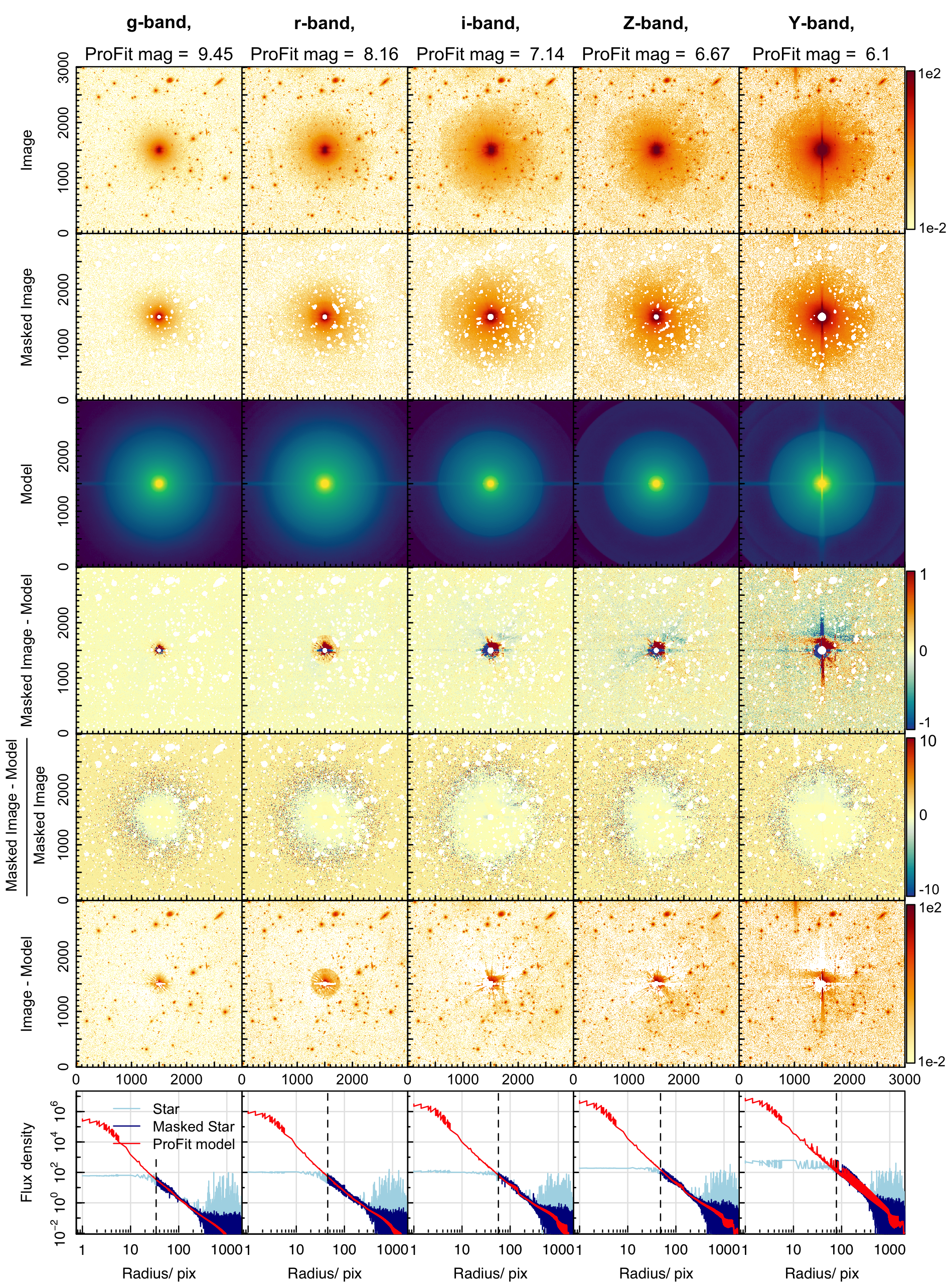}
    \caption{HSC-SSP PDR3 PSF application to model a real HSC star image, where each column represents each of the HSC $\textit{g,r,i,Z,Y}$ filters. The first row shows the image of the star up to an extent of R $\sim$ 1500 pix. The second row shows the masked version of the star image. The third row shows the \textsc{ProFit} model of the star, where we are using our HSC PSFs. The resulting fitted magnitudes per band are indicated in the headings of each column. In the next row, we show the result of subtracting the star \textsc{ProFit} model from the masked star image (2\textsuperscript{nd} row - 3\textsuperscript{rd} row), with a color scale between -1 and 1. The fifth row shows the subtraction of the masked image from the star model divided by the masked image. The sixth row shows the result of subtracting the star \textsc{ProFit} model from the star image (1\textsuperscript{st} row - 3\textsuperscript{rd} row), but in this case we use the same color scale as the 1\textsuperscript{st} row, mocking the star subtracted image with real flux values. The bottom panels show the radial distribution of the star in light blue, the radial distribution of the masked version of the star in dark blue, and the profile of the \textsc{ProFit} star model in red. The dashed black line indicates the radius up to where we mask the core of the star.}
    \label{PSF star application}
\end{figure*}

\subsection{Mock image application}
\label{mock}
In this application we generate two sets of mock images and analyse the impact of the $\textit{r}$-band HSC-SSP PDR3 PSF-scattered light on the faint end of the individual surface brightness distributions. Both sets consist of 100 mock images of 2000 $\times$ 2000 pixels.  

Each mock image in the first set simulates a different field of galaxies where the density of galaxies and their properties mimic a typical GAMA field (200 galaxies per image). We use \textsc{ProFit} to generate the properties of the galaxies following a Sérsic profile \citep{1963BAAA....6...41S}. The Sérsic profile template offers high flexibility in profile characteristics to generate a diverse range of galaxy types. Each Sérsic model is specified through the provision of eight parameters: position on the x-axis, position on the y-axis, magnitude, effective radius, Sérsic index, mayor-axis angle, minor to major axis ratio, and boxiness. These parameters are randomly sampled from a prior uniform distribution with reasonable limits. Table~\ref{Sersic} shows the range of values we provide to sample the parameters of each Sérsic model.

\begin{table}
\centering
\begin{tabular}{|c|c|}
 \textbf{Parameter} &\textbf{Allowed sampling region}                                                                                                                                                      \\ \hline
                     xcen   &       0 -- 2000                                      \\ \hline
ycen  &
0 -- 2000 
\\ \hline
                     mag   &       12 -- 22                                                                                                                                                               \\ \hline 
                      re   &       1 -- 100           
                      \\ \hline 
                       nser  &       0.5 -- 8        
                      \\ \hline 
                      ang   &       0 -- 180           \\ \hline 
                      axrat   &       0.5 -- 1           \\ \hline 
                      box   &       --0.3 -- 0.3           
\end{tabular}
\caption{Parameters we use to generate each mock galaxy using the Sérsic profile template of \textsc{ProFit}. Each parameter is sampled from a uniform distribution constrained by the limits indicated in the second column. }
\label{Sersic}
\end{table}

The second set of mock images replicates the field of galaxies generated in the first set, with identical spatial positions. The difference is that each mock image of the second set additionally contains a stellar distribution of 116 randomly placed stars. We use TRIdimensional modeL of thE GALaxy (TRILEGAL\footnote[7] {http://stev.oapd.inaf.it/cgi-bin/trilegal}, \citealt{2005A&A...436..895G}) to simulate our stellar number counts and HSC $\textit{r}$-band magnitude distribution. The TRILEGAL predicted number of counts per square degree is 13291, which is equivalent to 116 stars in our mock field of 0.0087 deg$^2$ (2000 $\times$ 2000 pixels). The HSC $\textit{r}$-band magnitude of each star is sampled from the TRILEGAL distribution, which varies between mag$_{\mathrm{r}}$ $\sim$ 6 -- 34.

We convolve each mock image of both sets with the $\textit{r}$-band HSC-SSP PDR3 PSF. In Fig.~\ref{convolution} we show a mock image from the second set before (left panel) and after (middle panel) the convolution with the PSF, where the pink circles indicate the position of the stars. The right panel shows the difference between the PSF-convolved version and the original mock image (2\textsuperscript{nd} panel - 1\textsuperscript{st} panel). Since the stars are simulated as point-like sources represented by single pixels, these are not noticeable in the left panel. However, we can appreciate the brightest point-like sources in the middle panel as the PSF has a larger spread effect on them.

In the first set, as there are no stars, we analyse the impact of the PSF only in extended sources. The upper panel of Fig.~\ref{trilegal} shows the 2D surface brightness grid of the difference in surface brightness ($\Delta_{\mathrm{SB}}$) between the 100 original mock images and the PSF-convolved versions from Set 1.  Dark blue SB bins indicate a higher number of counts, while pale blue bins correspond to a lower count. The majority of the pixels do not experience any SB difference when convolving with the HSC PSF, where there is a cloud of pale blue SB bins that suffer a difference of -- 2 $\lesssim$ $\Delta_{\mathrm{SB}}$ $\lesssim$ 4.

The lower panel of Fig.~\ref{trilegal} shows the 2D surface brightness grid of $\Delta_{\mathrm{SB}}$ for the second set of mock images, which additionally contain a stellar distribution in each field. We notice that a higher number of pixels suffer a SB increase in this case, with -- 2 $\lesssim$ $\Delta_{\mathrm{SB}}$ $\lesssim$ 9.

The PSF scatters light from the center of a source to the surrounding pixels, causing a decrease in the surface brightness of the central pixels ($\Delta_{\mathrm{SB}}$ $/<$ 0, blue pixels in the right panel of Fig.~\ref{convolution}) and an increase in the surface brightness on the outskirts ($\Delta_{\mathrm{SB}}$ $/>$ 0, red pixels in the right panel of Fig.~\ref{convolution}). The quantity of pixels that suffer a SB increase or decrease depends on the shape of the source. For a point-like source, only the central pixel has $\Delta_{\mathrm{SB}}$ $/<$ 0, as the stars are single bright pixels. However, a large surrounding area is affected by the PSF-spreading effect, causing an increase in the SB. We can see a similar effect in sharply profile galaxies. On the other hand, for smoother galaxy profiles, the relative number of pixels with $\Delta_{\mathrm{SB}}$ $/<$ 0 is higher in the center and the number of pixels with $\Delta_{\mathrm{SB}}$ $/>$ 0 is smaller on the outskirts.

In the lower panel of Fig.~\ref{trilegal}, as point-like sources are also contributing with spread flux to the faint end of the SB distribution, $\Delta_{\mathrm{SB}}$ goes up to 9. When estimations of low-surface brightness sources are required, this extra flux could be confused with real faint flux, leading to inaccurate measurements. A good star subtraction is therefore crucial to remove the contaminating light and correctly characterise LSB properties.

\begin{figure*}
\centering  
 \includegraphics[width=1\linewidth]{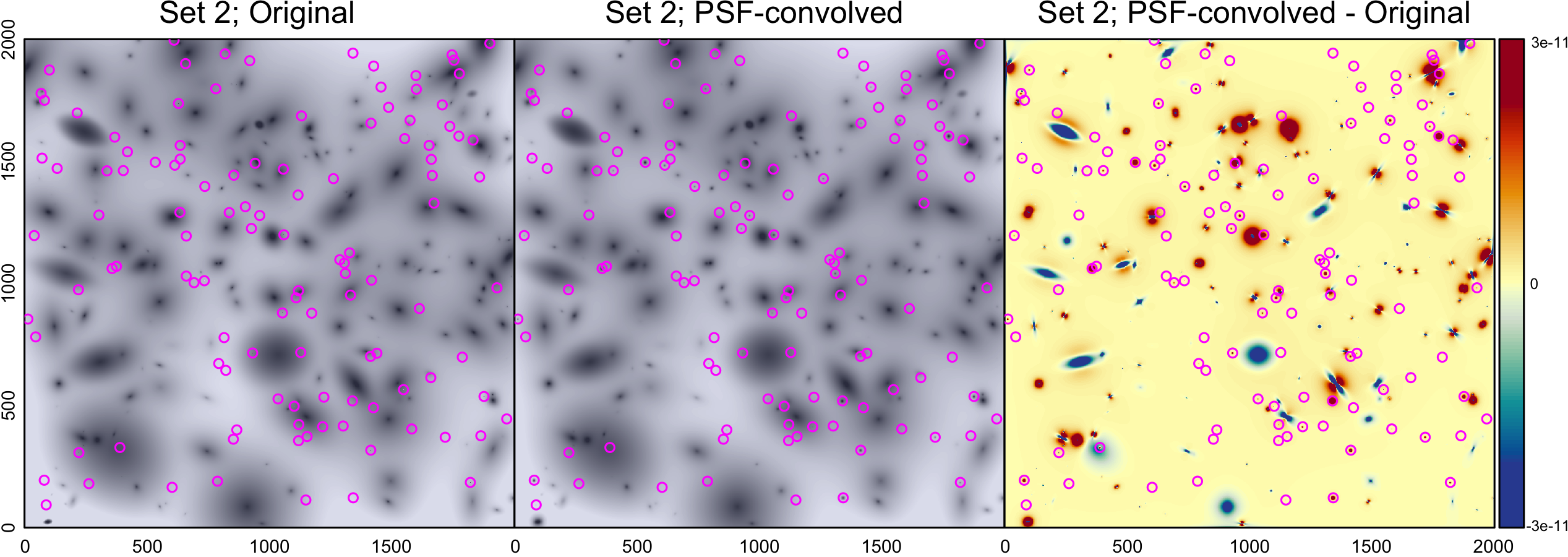}
    \caption{Mock image of a GAMA-like field of galaxies + TRILEGAL stellar distribution. The left panel shows the original mock image, the middle panel shows the $\textit{r}$-band HSC-PDR3 PSF-convolved version, and the right panel shows the difference between the PSF-convolved version and the original one. The pink circles indicate the position of the simulated point-like sources, with the PSF having the biggest effect on the brightest sources. In the right panel, positive pixels (red) contain higher values in the PSF-convolved image, and negative pixels (blue) contain smaller values. The PSF pushes flux toward the outskirts of any source, but the relative quantity of positive and negative pixels depends on the profile of the source. For point-like sources, only the central pixel is negative, and the area of surrounding positive pixels is larger. However, the relative negative area is larger in the center of smooth extended sources and the positive surrounding area is smaller.
    }
    \label{convolution}
\end{figure*}

Even though there is a large cloud of bins with $\Delta_{\mathrm{SB}}$ $\neq$ 0, the majority of the pixels do not suffer any change in their SB. We can observe this fact with the orange lines in Fig.~\ref{trilegal}, where the solid one indicates the median of the sample and the dashed lines indicate the 95$^{\mathrm{th}}$ and 99$^{\mathrm{th}}$ percentiles. This is because the Hyper Suprime-Cam is highly optimized for low surface brightness imaging and has a well-behaved PSF with low power distributed in the far wings.


\begin{figure}
\centering  
 \includegraphics[width=1\linewidth]{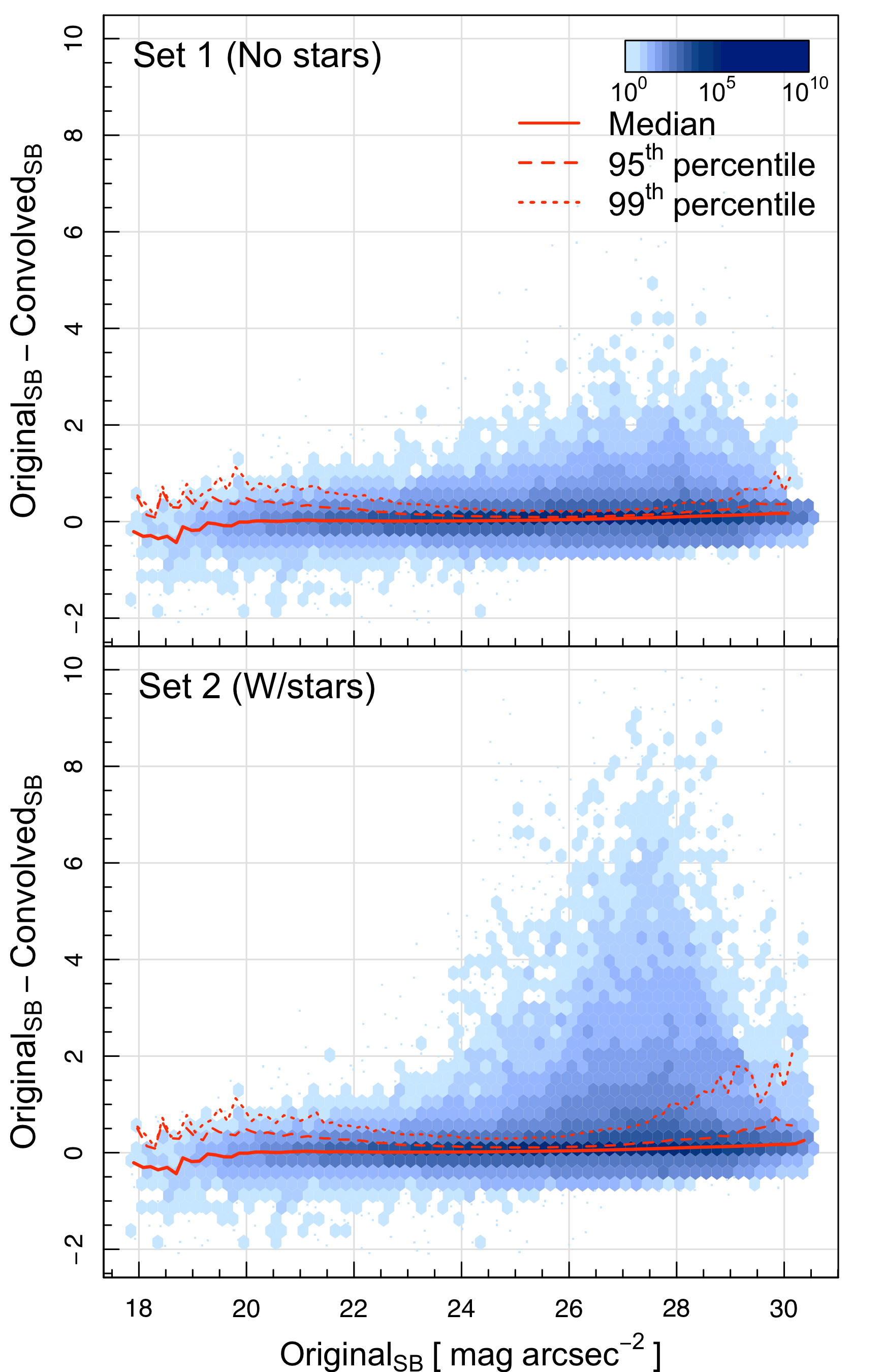}
    \caption{Difference of surface brightness between each mock image and its PSF-convolved version as a function of surface brightness for the two sets of mock images. The upper panel is generated using 100 mock images that only contain galaxies (Set 1), whereas the bottom panel is generated with 100 mock images that contain the same galaxies + stellar distributions (Set 2). The orange solid line indicates the median of the distribution and the two orange dashed lines indicate the 95$^{\mathrm{th}}$ and 99$^{\mathrm{th}}$ percentiles. The PSF has a major spread effect on the second set, as the stars contribute with extra flux to the neighbour pixels.
    }
    \label{trilegal}
\end{figure}

\section{Conclusions}
In this paper, we make public our new HSC-SSP PDR3 empirical PSF models for all $\textit{g,r,i,Z,Y}$ HSC filters. Using stacks of star images within a wide brightness range and following the method outlined by \cite{2020MNRAS.491.5317I}, we characterise the PSFs up to an extent of R $\sim$ 5.6 arcmin. These extended models are appropriate for the intermediate-stage data where the final local sky subtraction has not been applied. At this state, the data processing pipeline preserves well the extended wings of bright objects and is especially stored for users interested in low surface brightness science. This is, our new HSC-SSP PDR3s models will have a significant impact on the study of LSB structures by assisting the HSC community in recovering the spread light from both point and extended sources, as well as removing the contaminating light at large angles.

We present two examples showing the performance of our PSFs in scientific applications. In the first application, we use the HSC PDR3 PSFs to fit a 2-D model of a HSC star image in each filter. We demonstrate that our PSF models effectively make a good characterisation of these star images with some expected residuals. These subtraction residuals come from the small differences between each individual star and the averaged PSF model. In the second application, we generate two sets of mock images and analyse the impact of the \textit{r}-band HSC-SSP PDR3 PSF-scattered light on the surface brightness distributions. We see a minor effect on the first set of mock images which only contain field galaxies. However, the PSF-scattered effect is bigger when we analyse the second set of mock images which contain the same distributions of galaxies plus stellar distributions. Additional flux could lead to incorrect measurements when working with LSB sources if a proper star subtraction is not applied to the image. This second application consists of the first step of our ongoing work on studying the intra-halo light component in groups and clusters of galaxies using HSC observations.

The 2-D HSC PDR3 PSFs models are available as FITS files at \,\url{https://github.com/luciagarate/HSC\_PSFs}, as well as the scripts to reproduce the construction process.

\label{conclusions}

\section*{Acknowledgements}

LPGN, ASGR, and SB acknowledge support from the ARC Future Fellowship scheme (FT200100375). LJMD acknowledges support from the ARC Future Fellowship scheme (FT200100055).

This paper is based on data collected at the Subaru Telescope and retrieved from the HSC data archive system, which is operated by the Subaru Telescope and Astronomy Data Center (ADC) at NAOJ. Data analysis was in part carried out with the cooperation of Center for Computational Astrophysics (CfCA), NAOJ. We are honored and grateful for the opportunity of observing the Universe from Maunakea, which has the cultural, historical, and natural significance in Hawaii. The Hyper Suprime-Cam (HSC) collaboration includes the astronomical communities of Japan and Taiwan, and Princeton University. The HSC instrumentation and software were developed by the National Astronomical Observatory of Japan (NAOJ), the Kavli Institute for the Physics and Mathematics of the Universe (Kavli IPMU), the University of Tokyo, the High Energy Accelerator Research Organization (KEK), the Academia Sinica Institute for Astronomy and Astrophysics in Taiwan (ASIAA), and Princeton University. Funding was contributed by the FIRST program from the Japanese Cabinet Office, the Ministry of Education, Culture, Sports, Science and Technology (MEXT), the Japan Society for the Promotion of Science (JSPS), Japan Science and Technology Agency (JST), the Toray Science Foundation, NAOJ, Kavli IPMU, KEK, ASIAA, and Princeton University. This paper makes use of software developed for the Large Synoptic Survey Telescope. We thank the LSST Project for making their code available as free software at \,\url{http://dm.lsst.org}.

All of the work presented here was made possible by the free and open R software environment \citep{R-core}. All figures in this paper were made using the R \textsc{magicaxis} package \citep{2016ascl.soft04004R}. This work also makes use of the \textsc{celestial} package \citep{2016ascl.soft02011R}.


\section*{Data Availability}
The HSC-SSP PDR3 PSFs FITS files and construction scripts are available at \,\url{https://github.com/luciagarate/HSC\_PSFs}. The images are from the Hyper Suprime-Cam Public Data Release 3 (HSC-PDR3; \citealt{2022PASJ...74..247A}). 




\bibliographystyle{mnras}
\bibliography{example} 




\appendix

\section{HSC-SSP PDR3 PSFs in the resting bands}
\label{appendix}

In Fig.~\ref{PSF all}, we present the HSC-SSP PDR3 $\textit{g,r,i,Z,Y}$-band PSFs up to an extent of R $\sim$ 5.6 arcmin (2000 pixels). The bleeding (horizontal feature caused by CCD saturation), halo (extending smooth halo around the centre of the PSF which size depends on the brightness of the star), and ghost (produced by internal reflections in the HSC and visible in the outer stack) artefacts previously mentioned for the \textit{g}-band, are also present in the rest of the filters. Three ghosts are visible across the five bands: the inner one at a radial distance of R $\sim$ 170 arcsec, the middle one at R $\sim$ 270 arcsec, and the outermost ghost one located at R $\sim$ 330 arcsec. Additional but fainter ghost reflections are present in the $\textit{g,r,i}$-bands. The last observable artefact is the channel-stop or vertical feature, caused by a diffraction pattern in the CCD at long wavelengths ($\sim$ 1$\mu$$m$) and only present in the $\textit{Y}$-band. In this filter, all the stacks (outer, middle, inner, and core) are affected by the channel-stop. 

\begin{figure*}
\centering  
 \includegraphics[width=1\linewidth]{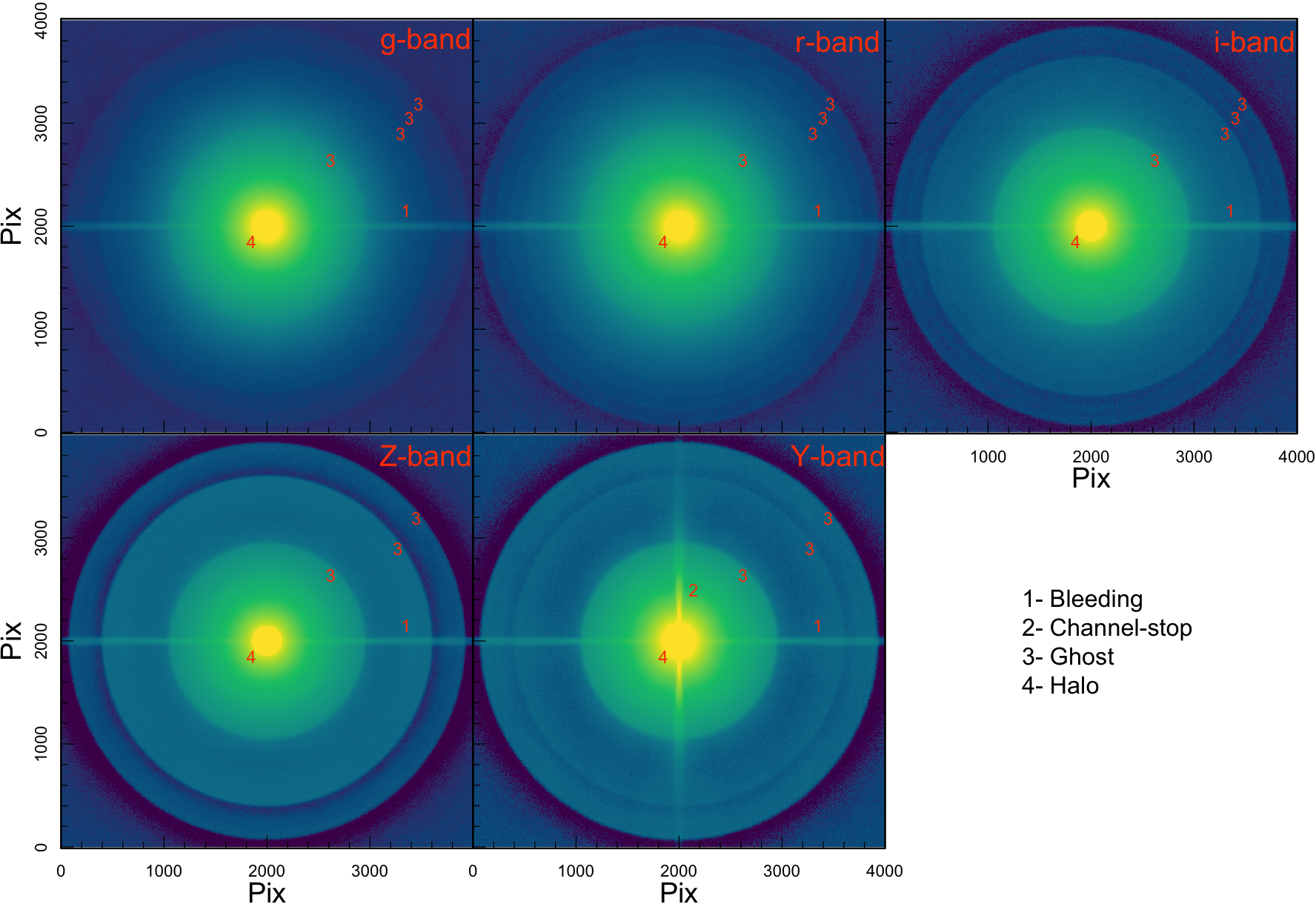}
    \caption{HSC-SSP PDR3 PSFs in the \textit{g,r,i,Z} and \textit{Y} filters up to an extent of R $\sim$ 5.6 arcmin (2000 pixels), where the optical and detector artefacts are being pointed in each panel. Bleeding (1) is the horizontal feature present in all the HSC filters, channel-stop (2) is the vertical feature (perpendicular to the bleeding direction) only present in the $\textit{Y}$-band, the edges of the ghosts features (3) are located at R $\sim$ 170 arcsec,  R $\sim$ 270 arcsec, and R $\sim$ 330 arcsec in the all the filters, with additional but fainter ghosts in the $\textit{g,r,i}$-bands. The smooth halo feature (4) is around the centre of each PSF. Any remaining negative pixels have been mapped to be at the bottom of the color scale.}
    \label{PSF all}
\end{figure*}

We finally show the radial distributions of all the HSC-SSP PDR3 in Fig.~\ref{PSF all profiles 1image}, where each color represents one filter. The spikes in the $\textit{Y}$-band correspond to the channel-stop higher flux values. There is a pair of horizontal and vertical dashed lines matched to each band, where the horizontal one indicates the maximum flux value of the PSF divided by two ($\mathrm{f_{max}}$/2), and the vertical line indicates the FWHM of the PSF also divided by two. By making an approximation across the five bands, the mean FWHM of the HSC-SSP PDR3 PSFs is 0.68 arcsec ($\sim$ 4 pixels). 

\begin{figure}
\centering  
 \includegraphics[width=1\linewidth]{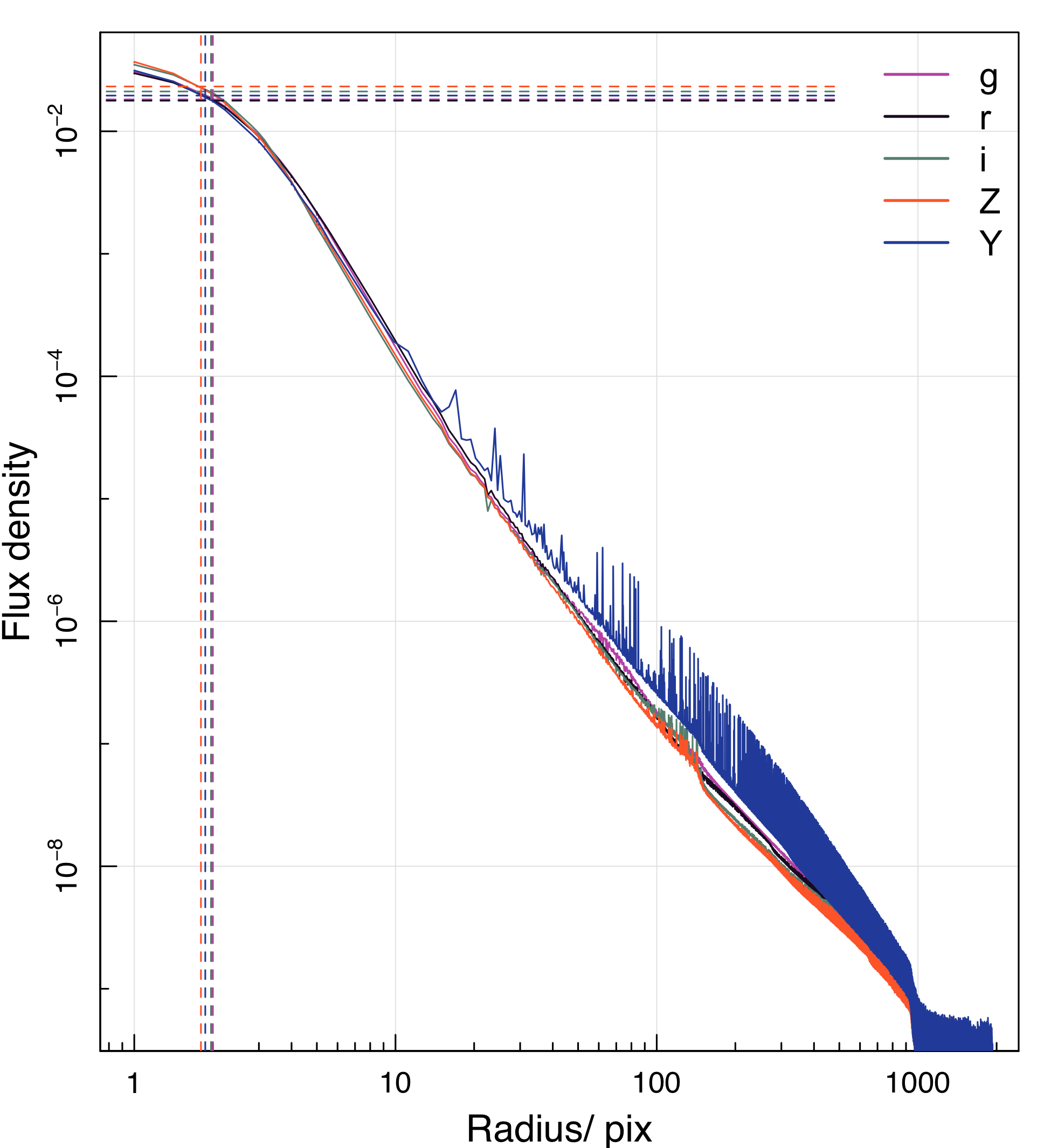}
    \caption{Radial HSC-SSP PDR3 PSF distributions in the five bands. The average FWHM is approximately 0.68 arcsec. The spikes of the $\textit{Y}$-band PSF correspond to the higher flux values of the vertical feature (or channel-stop) only present in this filter.}
    \label{PSF all profiles 1image}
\end{figure}

\section{Symmetric vs. Asymmetric PSF}
\label{comparison symm}
To quantify the disparity between the symmetric and the asymmetric PSFs, we analyse the differences of each relative to a model. This model is derived by fitting and smoothing the symmetric \textit{g}-band HSC-PDR3 PSF, but we could have selected another model, as the idea here is to see the comparison between different quantities (symmetric and asymmetric PSF) versus the same fixed quantity (model). In particular, we use the percentage error:
\begin{equation}    
\\
\\
\\
 \frac{\mathrm{PSF_{data} - Final_{model}}}{\mathrm{Final_{model}}}, \\
\end{equation}
\\
\noindent where $\mathrm{PSF_{data}}$ represents the constructed PSF and $\mathrm{Final_{model}}$ is a smooth model of the symmetric \textit{g}-band HSC-PDR3 PSF.

In Fig.~\ref{assym} we show the comparison between the \textit{g}-band HSC-PDR3 PSF before and after the symmetrisation process. The left panel shows the difference between both PSF images. The right panel shows the percentage error between the asymmetric PSF and the model of the symmetric PSF in pink and the percentage error between the symmetric PSF and the same model in magenta. As expected, the symmetrisation reduces the noise.

\begin{figure*}
\centering  
 \includegraphics[width=1\linewidth]{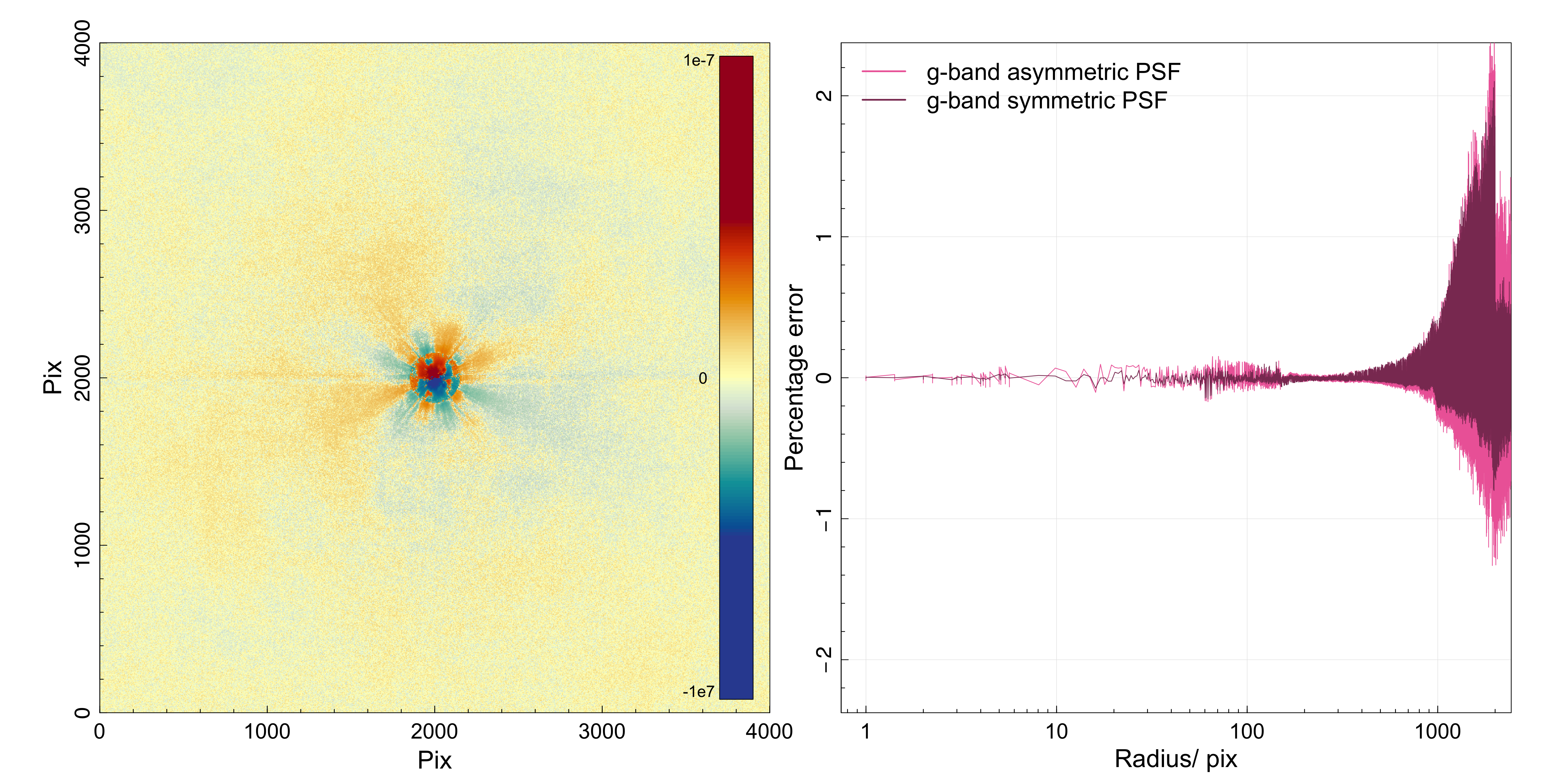}
    \caption{Left panel: difference between the \textit{g}-band symmetric HSC-PDR3 PSF and the \textit{g}-band asymmetric HSC-PDR3 PSF. Right panel: the pink curve shows the difference between the asymmetric PSF and a fixed model and the magenta curve shows the difference between the symmetric PSF and the same model, both in terms of percentage error.}
    \label{assym}
\end{figure*}


\bsp	
\label{lastpage}
\end{document}